\documentclass[preprint,showpacs,nofootinbib,preprintnumbers,amsmath,amssymb]{revtex4}
\usepackage{graphicx}
\usepackage{graphicx}
\usepackage{dcolumn}
\usepackage{bm}
\newcommand{\be}{\begin{eqnarray}}
\newcommand{\ee}{\end{eqnarray}}
\newcommand{\ave}[1]{\langle #1 \rangle}

 \newcommand{\gsim}{\mathrel{\hbox{\rlap{\lower.55ex \hbox {$\sim$}}
                   \kern-.3em \raise.4ex \hbox{$>$}}}}
\newcommand{\lsim}{\mathrel{\hbox{\rlap{\lower.55ex \hbox {$\sim$}}
                   \kern-.3em \raise.4ex \hbox{$<$}}}}

\begin{document}

\title{\textbf{{On the Charge Separation Effect\\ in Relativistic Heavy Ion Collisions} }}

\author{Jinfeng Liao $^1$, Volker Koch $^1$, and Adam Bzdak $^{1,\,2}$}

\affiliation{$^1$Nuclear Science Division, Lawrence Berkeley
National Laboratory, MS70R0319, 1 Cyclotron Road, Berkeley,
California 94720, USA.\\
$^2$Institute of Nuclear Physics, Polish Academy of Sciences,\\
Radzikowskiego 152, 31-342 Krakow, Poland.}

\begin{abstract}
In this paper, we discuss alternative means of measuring the possible
presence of local parity violation in  relativistic heavy ion
collisions.
We focus on the phenomenon of charge separation and introduce the
charged dipole vector $\hat{Q}^c_1$, which will measure the charge
separation on an event-by-event basis. Using Monte Carlo events, we
demonstrate the method and its
discriminating power. In particular we show that such an analysis will reveal
the strength of charge separation effect and its azimuthal correlation
with the reaction plane. We further show that our proposed
method may be able to distinguish between the actual charge separation
effect and effects due to certain two particle correlations.
The connection to present measurements based on particle
correlations is discussed.

\end{abstract}
\maketitle

\section{Introduction}

Topological objects in Quantum Chromodynamics(QCD) (and generally
in non-Abelian gauge theories) have attracted persistent
theoretical interests and are important in many aspects
\cite{'tHooft:1999au}. For example, instantons are known to be
responsible  for various properties of the QCD vacuum, such as
spontaneous breaking of chiral symmetry and the $U_A(1)$ anomaly
(see e.g. \cite{Schafer:1996wv}\cite{ES_book}). Magnetic
monopoles, on the other hand, are speculated to be present in the
QCD vacuum in a Bose-condensed form which then enforce the color
confinement, known as the dual superconductor scenario for QCD
confinement which is strongly supported by evidences from lattice
QCD (see e.g. \cite{Ripka:2003vv}\cite{Bali:1998de}).
Alternatively vortices are also believed to describe the
chromo-electric flux configuration (i.e. flux tube) between a
quark-anti-quark pair in the QCD vacuum which in turn gives rise
to the confining linear potential (see e.g. reviews in
\cite{Bali:1998de}\cite{Greensite:2003bk}). Some of these objects,
such as monopoles \cite{Liao:2006ry} and flux tubes
\cite{Liao:2007mj}, may also be important degrees of freedom in
the hot and deconfined QCD matter close to the transition
temperature $T_c$, and may be responsible for the observed
properties of the so called strongly coupled quark-gluon plasma
\cite{Shuryak:2008eq}. Certain phenomenological consequences of
such topological objects for relativistic heavy ion collisions
have been studied in \cite{Liao:2008dk}.

A particularly interesting suggestion by Kharzeev and
collaborators
\cite{Kharzeev:2004ey,Kharzeev:2007tn,Kharzeev:2007jp,Fukushima,Buividovich:2009wi,Kharzeev:2009fn}%
on the direct manifestation of effects from topological objects is
the possible occurrence  of $\cal P$- and $\cal CP$-odd (local)
domains due to the so-called sphaleron  transitions in the hot
dense QCD matter created in the relativistic heavy ion collisions.
In particular, the so called
Chiral Magnetic Effect(CME)\cite{Kharzeev:2007jp} predicts that in the
presences of the
strong external (electrodynamic) magnetic field  at the early
stage after a (non-central) collision sphaleron transitions induce a
separation of charges along the direction of the magnetic field.
Since the external magnetic field is
perpendicular to the reaction plane defined by the impact
parameter and the beam axis, one expects an out-of-plane charge
separation. As a
result positive charges are expected to preferentially go in one
(out-of-plane) direction and negative charges in the opposite
(out-of-plane) direction. In a given event, this charge separation
results in a momentum space electric dipole which  breaks
parity. However, the dipole moment will be, with equal probability,
parallel or anti-parallel to the magnetic field depending whether the Chiral
Magnetic Effect is caused by a sphaleron or anti-sphaleron
transition. Consequently, the expectation value of the dipole or,
more precisely, of the scalar product of the dipole and the
magnetic field, will vanish.

For the aforementioned reasons the CME will {\em not} give rise to a
non-vanishing  expectation value of a ${\cal P}$-odd observable.
However, the fact that
parity is broken event-by-event should be reflected in the
variance of a $\cal P$-odd observable, which, however is a $\cal
P$-even observable. Therefore, other, non-parity violating processes
may contribute which need to be well understood.

Very recently the STAR collaboration has announced the first
experimental evidence of a possible local parity violation
phenomenon at the Brookhaven's Relativistic Heavy Ion
Collider(RHIC) \cite{Star:2009uh}. STAR has  measured  the
differences between the in-plane-projected and
out-of-plane-projected 2-particle azimuthal correlations for both
same-charge pairs and opposite-charge pairs, as proposed by
Voloshin in \cite{Voloshin:2004vk}. The data  indeed show very
interesting charged-pair correlation patterns that depend on the
charges (same/opposite), reaction plane (in-plane/out-of-plane),
average $p_t$ of pairs, and colliding nuclei (Au/Cu). At first
sight, some features of the data appear to be consistent with what
has been expected from the local parity violation phenomenon.
However, as shown in Ref.\cite{Bzdak:2009fc}, certain aspects of
the data appear to be puzzling at present: contrary to
expectations from the Chiral Magnetic effect, the same-sign pairs
show a negative in-plane instead of a positive out-of-plane
correlation. Consequently, an interpretation of the data in terms
of  local-parity-violation would require a nearly exact
cancellation for all centralities of correlations due to the
Chiral Magnetic Effect and those due to ordinary correlations. In addition,
further studies have proposed alternative contributions to the
observed signals \cite{Wang:2009kd,Pratt:2010gy}, and
various possible consequences related to the local parity
violation \cite{Millo:2009ar} and the Chiral Magnetic
Effect \cite{Fukushima:2010vw,Basar:2010zd} have been recently
discussed.  At present, therefore, the existence of local parity
violation in heavy ion collisions has not yet been definitively
established and  further detailed and more differential analysis
of the current observable as well as the development of
alternative observables are necessary.

The purpose of this paper is to propose and study in detail an
alternative observable which specifically measures the charge
separation predicted by the CME. To this end we generally
investigate the charge separation effect as an intrinsic
charge-dependent particle azimuthal distribution, and propose a
way of measuring the magnitude of the charge separation effect and
its orientation relative to the reaction plane. The orientation,
if experimentally measured, will be essential for an evaluation of
the proposed local parity violation and CME interpretation.

In the
following, we will first introduce an intrinsic charge-dependent
distribution representing the charge separation effect. We will
then propose the charged dipole vector $\hat{Q}^c_1$ analysis
method and demonstrate its discriminating power by applying it to
Monte Carlo events. We will also present a detailed study of the
impact of certain two-particle correlations which may contribute
to charge separation and cause ambiguity in the interpretation of
the present STAR data. We will show that even with in the presence
of these  background correlations the  measurement of
$\hat{Q}^c_1$ together with the correlations proposed by STAR may
be able to unravel the charge separation, its orientation and  the
various sources contributing to it.

\section{The Charge Separation Effect}
\label{sec:sect2}

\begin{figure}[b]
\begin{center}
\includegraphics[width=8cm]{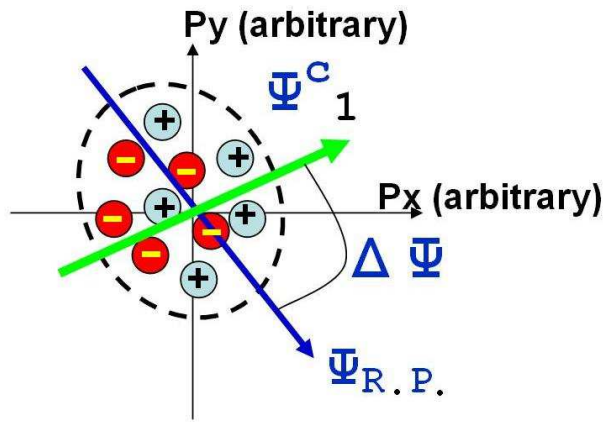}
\end{center}
\caption{A schematic demonstration of the proposed simultaneous
analysis of $\hat{Q}^c_1$ and $\hat{Q}_2$ vectors in the same
event. } \label{fig_demo}
\end{figure}

Let us begin by specifying  what we mean by the charge separation
effect in this paper. Consider the distribution of final state
hadrons in the transverse momentum space as schematically shown in
the Fig.\ref{fig_demo}. If the ``center'' of the positive charges
happens to be different from that of the negative charges, then
there is a separation between two types of charges which may be
quantified by an ``electric dipole moment'' in the transverse
momentum space. Such a separation may arise either simply from
statistical fluctuations or may be due to specific dynamics, such
as the Chiral Magnetic Effect. The later case is best discussed in
the {\em intrinsic} frame, i.e. the frame defined by the direction
of the reaction plane $\Psi_{R.P.}$. There, we can define an {\em
intrinsic}
 charge-dependent single
particle azimuthal distribution, which, besides a possible
momentum-space electric dipole moment, also allows for the
presence of elliptic flow.
\begin{eqnarray}\label{eqn_dist}
f_\chi\,(\phi,q) \propto 1+2\, v_2\cos(2\phi-2\Psi_{R.P.})+2\, q
\, \chi\, d_1\cos(\phi-\Psi_{C.S.})
\end{eqnarray}
Here $q$ and $\phi$ represent the charge and the azimuthal angle
of a particle, respectively. The parameters $v_2$ and $d_1$ quantify
the elliptic flow and the charge separation effect, while
$\Psi_{C.S.}$ specifies the azimuthal orientation of the
electric-dipole (see Fig.\ref{fig_demo})
and $\Psi_{R.P}$ the direction of the reaction plane.
It is important to
notice that an additional random variable $\chi=\pm 1$ is
introduced. This  accounts for the fact that in a given event we
may have sphaleron or anti-sphaleron transitions resulting in
charge separation parallel or anti-parallel to the magnetic field.
Consequently the sampling over all events with a given reaction
plane angle, $\Psi_{R.P}$, corresponds to averaging  the
intrinsic distribution $f_\chi$ over $\chi$, namely
$f=<f_\chi>_\chi\propto 1+2\, v_2\cos(2\phi-2\Psi_{R.P.})$.
Physically speaking this means that the charge separation (or electric
dipole, being $\cal P$-odd) flips and averages out to zero, thus
causing the expectation value of any parity-odd operator to
vanish. However, since $\ave{\chi^2} =1$ the presence
 of an event-by-event electric dipole may be observable in the
{\em variance} of a parity-odd operator.

Provided an accurate identification of the reaction plane is
possible one may also redefine the angles by setting
$\Psi_{R.P.}=0$ and replacing $\Psi_{C.S.}$ by $\Delta\Psi_{C.S.} \equiv
\Psi_{C.S.}-\Psi_{R.P.}$ in Eq.(\ref{eqn_dist}). For $\Delta\Psi_{C.S.}
=\pi/2$ the charge separation term takes the form of $d
\sin(\phi)$ as expected from the Chiral Magnetic Effect
\cite{Kharzeev:2009fn,Star:2009uh,Voloshin:2004vk}.

For measurements related to heavy ion collisions one may
reasonably assume particle charges to be $|q|=1$ which is the case
for almost all charged particles e.g. charged pions and kaons,
protons, etc. We also emphasize that  the above distributions does
{\em not} contain a directed flow term for either type of charges.
Finally one may also consider a $p_t$-differential formulation of
the charge separation effect or charge separation effects
associated with higher harmonics in the azimuthal angle $\phi$.

Let us next discuss how the above defined  charge-dependent
intrinsic single-particle distribution contributes to the charged
particle correlations recently measured by the STAR collaboration
in \cite{Star:2009uh}. Before doing so we note that there will
likely be additional contributions from two- and multi-particle
correlations which we will not consider in this Section. The STAR
collaboration has measured charge dependent two- and
three-particle correlations \cite{Star:2009uh}. Specifically they
considered

(i) The two-particle correlation $<\cos(\phi_i-\phi_j)>$ for
same-charge pairs ($++/--$) and opposite-charge pairs ($+-$). The
contribution to this correlator due to the charge-depended
intrinsic single-particle distribution, Eq.(\ref{eqn_dist}) is:
\begin{eqnarray}
&&\label{eqn_s_ij} <\cos(\phi_i-\phi_j)>_{++/--}\, =d_1^2 \\
&&\label{eqn_o_ij} <\cos(\phi_i-\phi_j)>_{+-}\, =-d_1^2
\end{eqnarray}

(ii) The three-particle correlation
\mbox{$<\cos(\phi_i+\phi_j-2\phi_k)>$} for same-charge pairs
($i,j=++/--$) and opposite-charge pairs ($i,j=+-$) with the third
particle, denoted by index $k$, having any charge. The
contribution to these correlators due to the distribution,
Eq.(\ref{eqn_dist}) turns out to be
\begin{eqnarray}
&& <\cos(\phi_i+\phi_j-2\phi_k)>_{++/--,\, k-any}\, = v_2\, d_1^2
\cos(2\, \Delta\Psi_{C.S.}) \\
&& <\cos(\phi_i+\phi_j-2\phi_k)>_{+-,\, k-any}\, = -v_2\, d_1^2
\cos(2\, \Delta\Psi_{C.S.})
\end{eqnarray}
where  ``k-any'' indicates that  the charge of the 3-rd particle
may assume any value/sign. The STAR collaboration has demonstrated
\cite{Star:2009uh} that the above three particle correlator is
dominated by the reaction plane dependent two-particles
correlation function \mbox{$<cos(\phi_i+\phi_j-2\Psi_{R.P.})>$}
and within errors they have found that
\begin{eqnarray}
<\cos(\phi_i+\phi_j-2\phi_k)> = v_2
<cos(\phi_i+\phi_j-2\Psi_{R.P.})>
\end{eqnarray}
Based on the distribution Eq.(\ref{eqn_dist}) we find the same
relation between these correlation functions, since the
reaction-plane dependent two-particle correlation is given by
\begin{eqnarray} \label{eqn_s_corr}
&& <\cos(\phi_i+\phi_j-2\Psi_{R.P.})>_{++/--}\, =d_1^2 \cos(2\,
\Delta\Psi_{C.S.})
\end{eqnarray}
for same-charge pairs, and
\begin{eqnarray} \label{eqn_o_corr}
&& <\cos(\phi_i+\phi_j-2\Psi_{R.P.})>_{+-}\, =-d_1^2 \cos(2\,
\Delta\Psi_{C.S.})
\end{eqnarray}
for opposite-charge pairs.

The proposed Chiral Magnetic Effect corresponds to $\Delta \Psi_{C.S.}
=\frac{\pi}{2}$. In this case \be
<\cos(\phi_i+\phi_j-2\Psi_{R.P.})>_{++/--}\, &=& -\,d_1^2 < 0,
\\
<\cos(\phi_i-\phi_j)>_{++/--}\, &=& +\,d_1^2 > 0. \ee Thus, the
two correlation functions are expected to be equal in magnitude
with {\em opposite} sign. The STAR measurement, on the other hand
finds them approximately equal in magnitude but with the {\em
same} (negative) sign. This discrepancy, which is discussed in
detail in \cite{Bzdak:2009fc}, needs to be understood before any
definitive conclusions about a possible charge separation effect
can be drawn. One aspect is the effect of higher order
correlations, which we have so far ignored.

In the following Section we present an alternative observable
which is sensitive to a potential charge separation effect. As we
will discuss in detail, this observable will take into account
additional correlations beyond those considered by STAR and thus
may help to clarify the present situation.

\section{Measurement of Charge Separation by $\hat{Q}^c_1$ Analysis}

In this Section we will present a method to directly measure the
intrinsic charge-dependent distribution in Eq.(\ref{eqn_dist}). We
propose  to measure the charged dipole moment vector $\hat{Q}^c_1$
of the final-state hadron distribution in the transverse momentum
space. The magnitude $Q^c_1$ and azimuthal angle $\Psi^c_1$ of
this vector can be determined in a given event by the following:
\begin{eqnarray} \label{eqn_qc1_def}
Q^c_1 \cos \Psi^c_1 \equiv \sum_i q_i \cos\phi_i \nonumber \\
Q^c_1 \sin \Psi^c_1 \equiv \sum_i q_i \sin\phi_i
\end{eqnarray}
where the summation is over all charged particles in the
event with $q_i$ and $\phi_i$ the electric charge\footnote{As
already mentioned, one may reasonably assume $|q_i|=1$.} and
azimuthal angle of each particle. This method is in close
analogy to the $\hat{Q}_1$ and $\hat{Q}_2$ vector analysis used
for directed and elliptic flow (see e.g. \cite{Voloshin:2008dg}).
In particular the elliptic flow and the reaction plane orientation
can be determined via the $\hat{Q}_2$ vector (with magnitude $Q_2$
and azimuthal angle $\Psi_2$) as:
\begin{eqnarray} \label{eqn_q2_def}
Q_2 \cos 2\Psi_2 \equiv \sum_i \cos2\phi_i \nonumber \\
Q_2 \sin 2\Psi_2 \equiv \sum_i \sin2\phi_i
\end{eqnarray}
We emphasize that contrary to $\hat{Q}_2$ the charge dipole vector,
$\hat{Q}^c_1$, incorporates the {\em electric charge} $q_i$ of the
particles\footnote{ Mathematical
details regarding the observable $\hat{Q^c_1}$ can be found in the
Appendix.}.
 The angles $\Psi^c_1$ and $\Psi_2$ are determined
event-by-event from a finite number of particles  and are
not to be confused with the idealized expectations $\Psi_{C.S.}$
and $\Psi_{R.P.}$, although they should become identical in the limit
of infinite multiplicity.

With both the magnitude $Q^c_1$ and the azimuthal angle $\Psi^c_1$
determined on an event-by-event basis, important information about
the underlying physics can be revealed.\footnote{After the
completion of the present work, we became aware of the recent
PHENIX efforts \cite{Lacey} to measure the distribution of the
difference of $\sum \sin(\phi_i)$ between plus and minus charges.
This is related to the quantity $Q^c_1\sin\Psi^c_1$, which is
part of the full information that can be extracted by the $\hat
Q^c_1$ analysis.} In particular, the event-by-event distribution
of the magnitude $Q^c_1$  may indicate if  there is a physical
charge separation effect beyond pure statistical fluctuations.
Furthermore, the relative distribution of the angle $\Psi^c_1$
with respect to the reaction plane determined in the same event
 (e.g. via the $\hat{Q}_2$ analysis) can show to which extend the
charge separation is correlated
with the reaction plane with a specific angle $\Delta \Psi_{C.S}=\Psi_{C.S}-\Psi_{R.P.}$ (see
Fig.\ref{fig_demo}).

Next let us investigate the discriminating power
of the proposed $\hat{Q}^c_1$ analysis for a potential charge separation
effect. To this end  we employ Monte Carlo
sampling to generate an ensemble of final state
hadrons with equal numbers of $\pi^+$ and $\pi^-$. For
each particle we sample an azimuthal angle $\phi_i$ of its transverse
momentum $\vec p_t$. We do not sample the magnitude of the transverse momentum,
$p_t$, as the $\hat{Q}^c_1$ analysis defined in
Eq.(\ref{eqn_qc1_def}) involves only the angle $\phi_i$
\footnote{One may actually assign $p_t$-dependent weight factor in
the definition of $\hat{Q}^c_1$ to maximize manifestation of the
desired physical effect, as has been done in the $v_2$ analysis
(see e.g. \cite{Voloshin:2008dg}).}.

To mimic an event that may have elliptic flow and a possible
charge separation effect, we sample the azimuthal angles for
$N_+=200\,$ $\pi^+$ and $N_-=200$ $\pi^-\,$ according
to the intrinsic  distribution in Eq.(\ref{eqn_dist})
 \footnote{To be more
realistic one shall allow for fluctuations of the particle numbers,
$N_+$ and $N_-$, within a
selected multiplicity window. These additional sources of background
statistical fluctuations bring in negligible broadening of the
$\hat{Q^c_1}$ magnitude distribution curves in
Fig.\ref{fig_mag_dist} as we have verified numerically with up to $\pm
20\%$ fluctuations independently for  $N_+$ and $N_-$.}
\footnote{We emphasize again that in such sampling, {\em no}
directed flow effect will be generated and only pure statistic
fluctuations will contribute in the usual $\hat{Q}_1$ analysis.}.
The sampling parameter $v_2$  specifies
the magnitude of the elliptic flow  and $\Psi_{R.P.}$ the orientation of the
reaction plane, which is randomly chosen in each event.
The parameters $d_1$ and $\Psi_{C.S.}$
specify the magnitude of charge separation effect and the
orientation of the charged dipole, respectively. Finally for each
event we randomly pick the sign of the parameter $\chi=\pm 1$.

In order to investigate different physical situations we
consider a variety of cases as follows:\\

{\bf Case-Ia} with $d_1=0$ and $v_2=0$, and {\bf Case-Ib} with
$d_1=0$ but $v_2\ne 0$. In both cases the charge separation
may arise only from  statistical fluctuations with or without
the presence of elliptic flow;\\
{\bf Case-II} with $d_1\ne 0$ but $v_2=0$. Here we allow for an
explicit charge separation in addition to  statistical
fluctuations but do not consider any explicit elliptic flow;\\
{\bf Case-IIIa} with $d_1\ne 0$ and $v_2\ne 0$ but with the two
orientations $\Psi_{C.S.}$ and $\Psi_{R.P.}$ randomly assigned in
each event, representing a situation where both charge separation and
elliptic flow are
present but {\em not} correlated in the azimuth; \\
{\bf Case-IIIb} with $d_1\ne 0$ and $v_2\ne 0$ and with the two
orientations $\Psi_{C.S.}$ and $\Psi_{R.P.}$ always parallel in each event; \\
{\bf Case-IIIc} with $d_1\ne 0$ and $v_2\ne 0$ and with the two
orientations $\Psi_{C.S.}$ and $\Psi_{R.P.}$ always perpendicular
in each event; \\
{\bf Case-IIId} with $d_1\ne 0$ and $v_2\ne 0$ and with the two
orientations $\Psi_{C.S.}$ and $\Psi_{R.P.}$ always
$\frac{\pi}{4}$ apart in each
event.\\
A comparison between the various cases will allow us to test the
ability of the proposed analysis to discriminate between the various
physical scenarios.

For each of the above cases we sample 100 million events. For each
event we perform the  $\hat{Q}_2$ and $\hat{Q}^c_1$ analyses and
extract the values $Q_2,\Psi_2$ and $Q^c_1,\Psi^c_1$.
The resulting distributions of the magnitude
$Q^c_1$ and the angle $\Psi^c_1$ for the charged dipole vector
$\hat{Q}^c_1$ will be presented and discussed in the following subsections.

\subsection{The Magnitude Distribution}

\begin{figure}[t]
\begin{center}
\includegraphics[width=8cm]{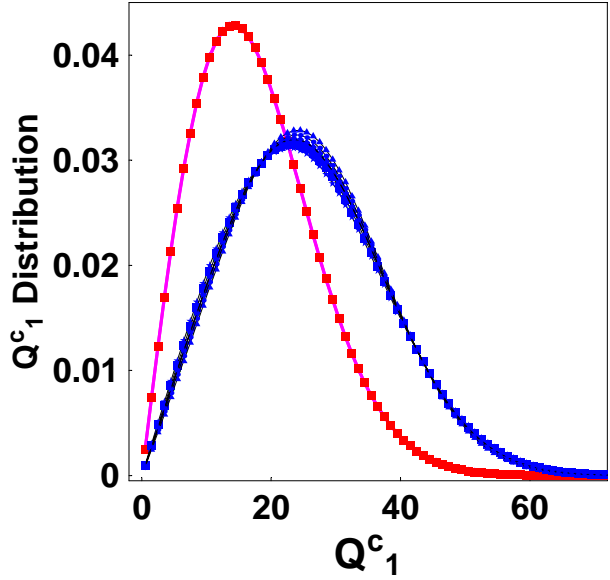}
\end{center}
\caption{The $Q^c_1$ magnitude distribution extracted from sampled
events for the various cases. The curves for Case-Ia (red) and for
Case-Ib with two values of $v_2$(magenta) appear on top of each
other at the left, while the curves for Case-II(green) and for
Case-IIIa,b,c,d with two values of $v_2$ for each(blue) appear on
top of each other at the right (see text for details).  }
\label{fig_mag_dist}
\end{figure}

The $Q^c_1$ magnitude distributions (normalized to  unity) for the
various aforementioned cases are shown in Fig.\ref{fig_mag_dist}.
We have chosen the following values for the sampling parameters:
For the cases with non-vanishing dipole moment, Case-II and
Case-IIIa,b,c,d, we have used $d_1=0.05$. For the cases with
non-vanishing elliptic flow, Case-Ib and Case-IIIa,b,c,d, we
allowed for two values of elliptic flow parameter $v_2=0.1$ and
$v_2=0.05$ in each case. Therefore, overall there are {\em twelve}
curves in the plot: interestingly, they all fall into only two
groups. One group includes all the curves corresponding to the
cases with $d_1=0$ (left set of curves in Fig.\ref{fig_mag_dist})
and the other corresponds to all cases with finite dipole moment,
$d_1=0.05$, irrespective of the presence and strength of elliptic
flow.

From these result we may conclude the following: (a) the $Q^c_1$
magnitude distribution is very sensitive to the physical charge
separation effect represented by $d_1$, but rather unaffected by
the elliptic flow $v_2$; (b) the left curve(s) in
Fig.\ref{fig_mag_dist} represents the pure
statistical fluctuation, and any deviation of the measured
distribution from this curve would contain information about the
physical charge separation effect. Experimentally the reference
curve from statistical fluctuation can be obtained by randomly
re-assigning  the charges of the particles in each event. We have verified with
our sampling, that the analysis after charge-reshuffling shifts all the
curves to the statistical ones, denoted as Case Ia and Ib
in Fig.\ref{fig_mag_dist}.

\subsection{The Angular Distribution}

We next turn to the angular distribution. Since any experiment
samples over all directions of the reaction plane the only
experimental accessible information is the relative angle, $\Delta
\Psi = \Psi_1^c-\Psi_2$, between the direction of the charge
dipole, $\Psi^c_1$, and that of the reaction plane, or more
precisely, the elliptic flow, $\Psi_2$.

We focus on the four cases Case-IIIa,b,c,d, where both the
elliptic flow and the charge separation effects are present. For
studies in the present subsection, we set $v_2=0.1$ and $d_1=0.05$
in the sampling. For each sampled event we then extract the angles
$\Psi^c_1$ and $\Psi_2$ from each event and calculate their
difference $\Delta\Psi=\Psi^c_1-\Psi_2$. Since the angle $\Psi_2$
(essentially the reaction plane orientation) is defined modulus
$\pi$, $\Delta\Psi$ is equivalent to $\Delta \Psi \pm \pi$ and we
may always transform $\Delta\Psi$ to be in the interval $ \Delta
\Psi \in [-\pi/2,\pi/2)$\footnote{In addition, the direction of
the
  electric dipole moment is only known modulo $\pi$ due to the random
  factor $\chi$ in Eq.(\ref{eqn_dist})
representing the equal likelihood of sphalerons and
  anti-sphalerons.}. Furthermore, what matters most for our
discussion is the distinction between the in-plane and the
out-of-plane orientations. Thus it is sufficient to know the
absolute value of $\Delta \Psi$, i.e. $ |\Delta \Psi|\in
[0,\pi/2]$.

\begin{figure}[t]
\begin{center}
\includegraphics[width=7.cm]{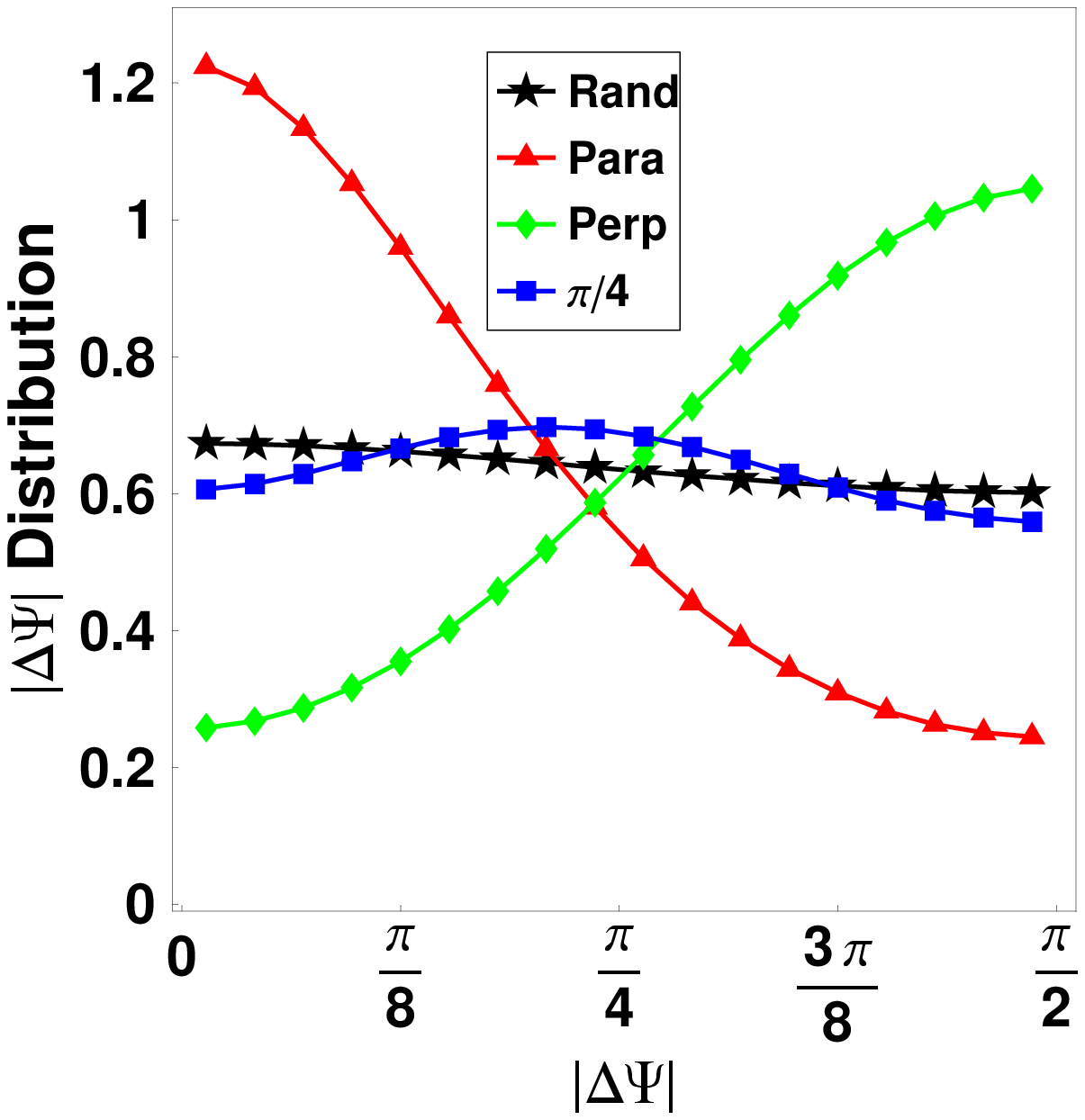}\hspace{0.5cm}
\includegraphics[width=7.5cm]{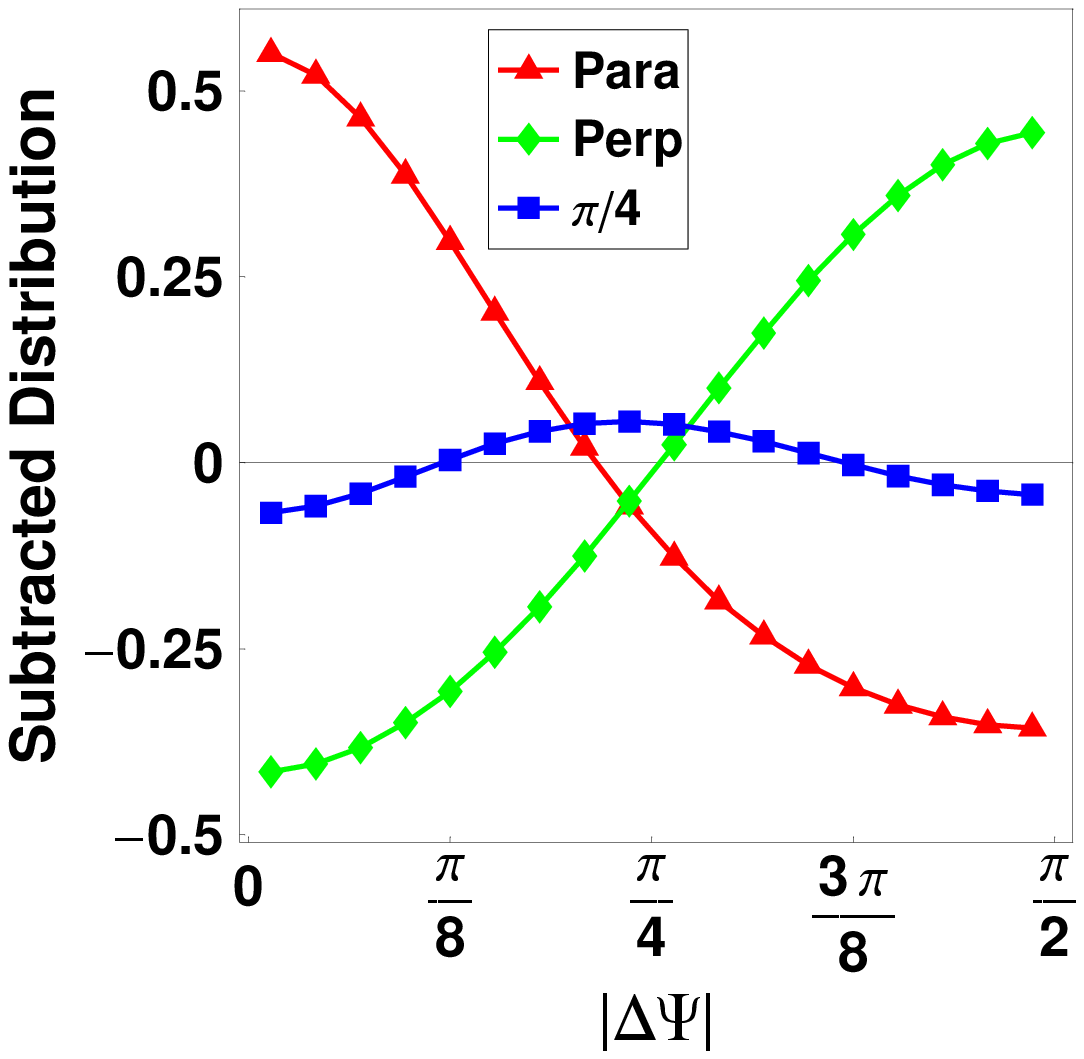}
\end{center}
\caption{(left) The distributions of the relative azimuthal angle
$|\Delta \Psi|$ between the charged dipole vector $\hat{Q}^c_1$
and the reaction plane for Case-IIIa(black star),b(red
triangle),c(green diamond),d(blue box). (right) The subtracted
distributions (i.e. the difference with respect to the Case-IIIa
distribution) for Case-IIIb(red triangle), Case-IIIc(green
diamond), and Case-IIId(blue box) respectively. (See text for more
details). } \label{fig_ang_dist}
\end{figure}

The resulting $|\Delta \Psi|$ distributions for the Cases
IIIa,b,c,d are shown in Fig.\ref{fig_ang_dist}(left). They clearly
exhibit distinctive patterns for the various cases. Out of the
four cases, the Case-IIIa serves as a background since in this
case the orientations of the elliptic flow and the dipole are {\em
uncorrelated}\footnote{We note that even for the Case-IIIa where
there is no a prior correlation in the sampling, the distribution
is not entirely flat but slightly favors a smaller angular
separation between $\Psi^c_1$ and $\Psi_2$. This is not surprising
as both $\hat{Q}^c_1$ and $\hat{Q}_2$ are determined from the same
set of particles in a given event. Given the probability
distribution, Eq.(\ref{eqn_dist}), it is clear that for any
angular sampling parameters $\Psi_{C.S.}$ and $\Psi_{R.P.}$ the
maximum of the probability distribution is located at certain
angle $\Psi_{max}$ (and $\Psi_{max}+\pi$ for the opposite charge)
between the two. As a result both angles, $\Psi^c_1$ and $\Psi_2$,
extracted from the sampled events tend to slightly align with the
maximal angle $\Psi_{max}$. In an actual data analysis, this
background Case-IIIa angular distribution can be well approximated
by the angular distribution from pure statistical background
obtainable via the previously mentioned charge-reshuffling, as we
have verified with our sampling.}. Any deviation from the
Case-IIIa $|\Delta \Psi|$ distribution indicates an azimuthal
correlation between the dipole and the elliptic flow.
 This background may be removed by subtracting the
results from Case-IIIa in our model. Such subtracted $|\Delta
\Psi|$ distributions for the Case-IIIb,c,d are shown in
Fig.\ref{fig_ang_dist}(right) for comparison.

The curve for Case-IIIb, where the orientations for the elliptic
flow and charge separation effects are taken to be parallel,
$\Psi_{C.S.}=\Psi_{R.P}$, shows a maximum at $|\Delta \Psi|=0$ (in
plane)  and a rapid decrease toward $|\Delta \Psi|=\pi/2$
(out-of-plane). Thus the dipole $\hat{Q}^c_1$ is predominantly
oriented in-plane. The curve for Case-IIIc, where the orientations
for the elliptic flow and charge separation effects are taken to
be  perpendicular, shows exactly the opposite. Finally the curve
for Case-IIId, where the orientations for the elliptic flow and
charge separation effects are sampled to be $\pi/4$ apart from
each other, features a maximum around $|\Delta \Psi|=\pi/4$ and a
decrease toward both ends. The effect, however is not as prominent
as in the previous cases and it may be more difficult to separate
it from the background Case-IIIa. For these scenarios  we have
also varied values of both $v_2$ and $d_1$ in our sampling and
found the patterns to be qualitatively the same (see Section
\ref{sec:disentangle} for details).

From these studies, one can conclude that by simultaneously
measuring the orientations of $\hat{Q}^c_1$ and $\hat{Q}_2$ and
examining their angular difference distribution, the azimuthal
correlation between the two physical effects may be extracted.

\subsection{An Alternative Measurement of Relative Orientation}

While measuring the relative orientation distribution as discussed
in the previous subsection contains the ``full'' information, it
is experimentally rather demanding as it requires determination of
reaction plane in each event. There is an alternative way to
determine whether the charged dipole $\hat{Q}^c_1$ is closer to the
in-plane or out-of-plane direction without the need of reaction
plane. The idea is to measure  the correlation function
$\ave{\cos(2\Delta \Psi)}=\ave{\cos(2\Psi_{C.S.}-2\Psi_{R.P.})}$,\footnote{A
similar correlation
between the directed flow and elliptic flow was proposed by
Poskanzer and Voloshin in \cite{Poskanzer:1998yz} to determine
their relative orientation.}.
which in a sense represents the ``elliptic anisotropy''
of $\hat{Q}^c_1$. By using the definitions in
Eqs.(\ref{eqn_qc1_def},\ref{eqn_q2_def}), one  obtains
\begin{eqnarray} \label{eqn_cos_2dPsi}
\ave{\cos(2\Delta \Psi)} = {\bigg <} \frac{N_{ch}   + 2\,
\{i,j\}^c_1 + \{i,j\}_2 +
\{i,j;k\}^c}{\left[N_{ch}+\{i,j\}^c_1\right]\cdot
\left[N_{ch}+\{i,j\}_2\right]^{1/2}} {\bigg >}
\end{eqnarray}
with the two- and three- particle correlations defined as
\begin{eqnarray} \label{eqn_correlations}
\{i,j\}^c_1 &&\equiv \sum_{i\ne j}q_i q_j \cos(\phi_i-\phi_j)
\nonumber \\%
\{i,j\}_2 &&\equiv \sum_{i\ne j}\cos2(\phi_i-\phi_j) \nonumber \\%
\{i,j;k\}^c &&\equiv \sum_{i\ne j \ne k} q_i q_j
\cos(\phi_i+\phi_j-2\phi_k)
\end{eqnarray}
(see the Appendix A for details).

If the observable $\ave{\cos(2\Delta \Psi)}$ is measured to be
unambiguously negative/positive then the charge separation is
closer to out-of-plane/in-plane. If it is however consistent with
zero within errors, then the situation is unclear. An ideal
Chiral Magnetic Effect without any other effect and without
statistical fluctuations would predict
$\ave{\cos(2\Delta \Psi)}=-1$.  As will be discussed in the next
Section, however, statistical fluctuations and additional
two-particle correlations may lead to sizable corrections.

Let us finally point out that the correlation function,
Eq.(\ref{eqn_cos_2dPsi}), in principle involves particle
correlations to all orders, since it represents an average of a
ratio. In addition, the correlators appearing in the denominator,
in particular $\{i,j\}_2$, is of the same magnitude as $N_{ch}$,
which makes an expansion of the denominator unreliable.
Consequently, the correlation function, Eq.(\ref{eqn_cos_2dPsi}),
has to be measured in its entirety and represents a different
measurement than that of the individual terms, which has been
already carried out by the STAR collaboration. A dedicated
measurement is thus important and urgently called for.

\section{Disentangling Charge Separation and Two-Particle
  Correlations}
\label{sec:disentangle}

In this Section we focus on the influence of certain two-particle
correlations on both the observables proposed in the present paper
and the correlations measured by STAR. As has been shown in
\cite{Bzdak:2009fc}, the present STAR data indicates the existence
of at least two types of correlations (beyond the possible
correlations predicted by the Chiral Magnetic Effect): same-charge
pair back-to-back correlations (mostly in-plane), and
opposite-charge pair same-side correlations (about equally
in-plane and out-of-plane). It is therefore important to study the
robustness of various observables against such ``background
correlations'', and in particular to evaluate the prospects of
disentangling a physical dipole charge separation from those
background correlations. Physically speaking, there are a number
of potential sources that may induce such two-particle
correlations, such as clusters, momentum and/or local charge
conservation \cite{Wang:2009kd}\cite{Pratt:2010gy}, or even chiral
magnetic spiral effect \cite{Basar:2010zd}.

In order to study correlation effects we introduce correlations on
top of our previous sampling based on single particle distribution
in Eq.(\ref{eqn_dist}). In particular we test the two types
indicated by data, i.e.: (a) same-charge pair back-to-back
correlation ({\bf SCBB}), which we implement by randomly selecting
a small fraction of same-charge pairs and sampling them according
to $|\phi_1-\phi_2|> \pi/2$;  (b) opposite-charge pair same-side
correlation ({\bf OCSS}), which we implement by randomly selecting
a small fraction of opposite-charge pairs and sampling them
according to $|\phi_1-\phi_2|< \pi/2$. At the same time we ensure
the single particle distribution remains unchanged.\footnote{The precise procedure we used here is the following: when sampling each particle (say a plus charge) we trigger the correlation with probability $0.4$ and once triggered we correlate it with the preceding plus/minus charge for SCBB/OCSS correlations by sampling its angle according to both the single particle distribution and the constraint from correlation. We have studied different implementations (which essentially imply different approximations to the desired correlations) and found that our conclusions are not changed. } In the
following we discuss various situations with different
choices for the strength of the
 dipole charge separation and the correlations. For all studies in
 this Section we set $v_2=0.1$ and generate 10 million events for each case.

\subsection{Test-I}

In Test-I, we consider a setting with relatively ``large''
out-of-plane dipole and small
correlations and study four cases:\\
{\bf Case-IIIc} (previously studied) with dipole $d_1=0.05$
and $|\Delta\Psi|=\pi/2$ in the sampling;\\
{\bf Case-IIIc-($\alpha$)} with the same dipole as in Case-IIIc, plus SCBB %
correlation at a fraction $f_{SCBB}= 0.4\%$ of all same-charge pairs   ;\\
{\bf Case-IIIc-($\beta$)} with the same dipole as in Case-IIIc, plus OCSS %
correlation at a fraction $f_{OCSS}= 0.4\%$ of all opposite-charge pairs ;\\
{\bf Case-IIIc-($\gamma$)} with the same dipole as in Case-IIIc,
plus both SCBB  and OCSS correlations at the same fraction of
$f_{SCBB} = f_{OCSS}= 0.4\%$  .

\begin{figure}[!h]
\begin{center}
\includegraphics[width=6.8cm]{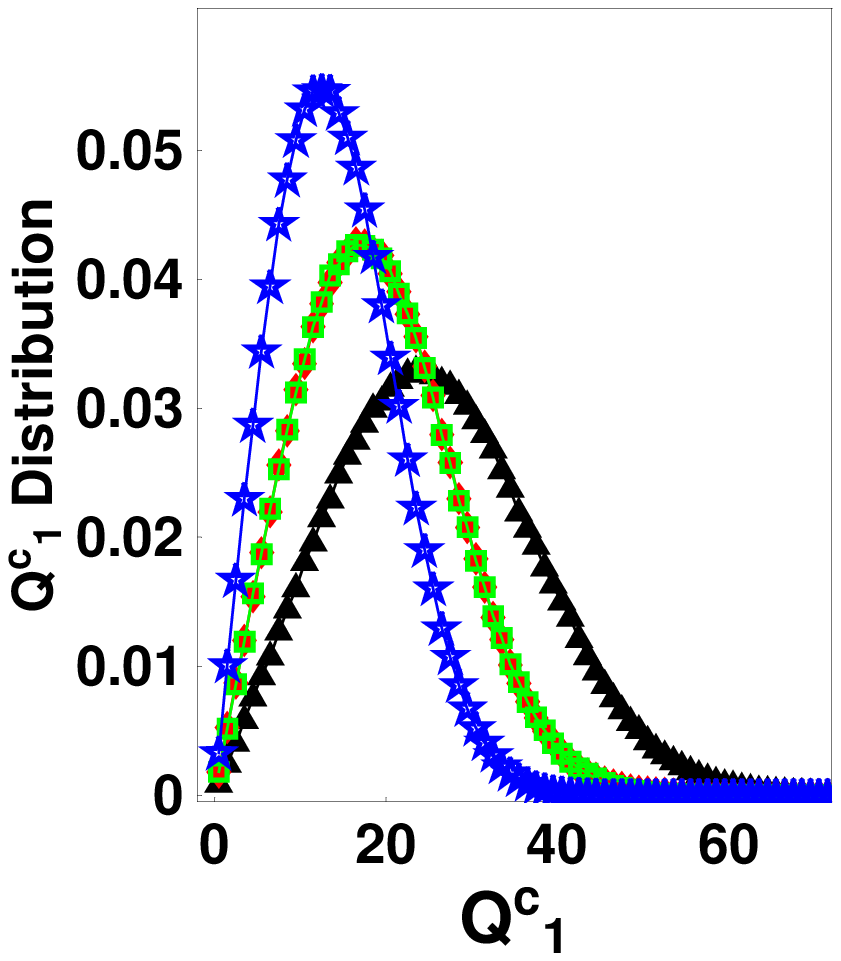}\hspace{0.75cm}
\includegraphics[width=6.5cm]{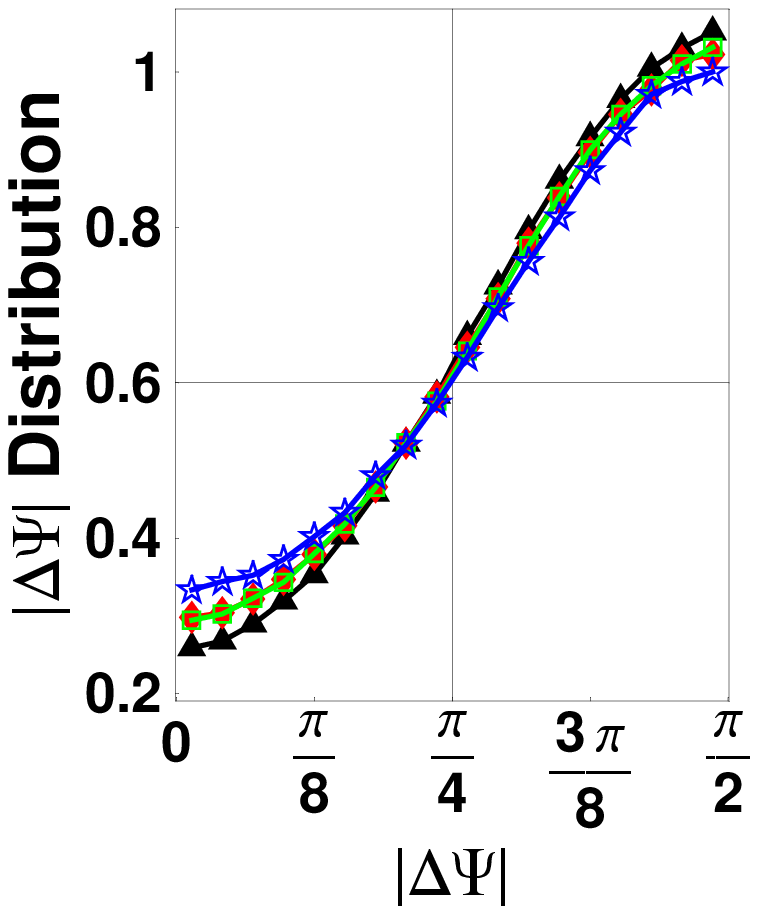}
\end{center}
\caption{The $Q^c_1$ (left) and $|\Delta \Psi|$  (right)
distributions for the four cases in Test-I. The triangle(black),
diamond(red), box(green), and star(blue) symbols are for
Case-IIIc, IIIc-($\alpha$), IIIc-($\beta$), and IIIc-($\gamma$)
respectively (see text for more details). } \label{fig_perp_corr}
\end{figure}

The results for $Q^c_1$ and $|\Delta \Psi|$ distributions are
shown in Fig.\ref{fig_perp_corr} and the various measured
observables \footnote{The definitions and interrelations of these
observables can be found in the Appendix.} are reported in
Table.\ref{table_1}. From these
results one can see that:\\
(1) Both types of correlations shift the $Q^c_1$ distribution towards
smaller values,  i.e. they {\em
suppress} the charge separation;\\
(2) The $|\Delta \Psi|$ distribution, on the other hand, is quite robust in
the present setting and and maintains a clear pattern
characteristic of an out-of-plane charge separation in all four
cases. However, the correlations lead to a slight flattening of the
distributions.\\
(3) Correspondingly, the observable $\left<\cos(2\,
\Delta\Psi)\right>$ remains negative for all cases, with its
absolute value slightly reduced by both types of correlations;\\
(4) The observable $\left<\cos(\phi_i+\phi_j-2\phi_k)\right>$ remains negative
for same-charge pairs and positive for opposite-charge pairs;\\
(5) The observable $\left<\cos(\phi_i-\phi_j)\right>$ has
completely different sign patterns in the four cases, and appears to
be a sensitive diagnostic for different types of correlations; \\
(6) The Case-IIIc-($\gamma$) (see last column in
Table.\ref{table_1}), where the dipole coexists with both SCBB and
OCSS correlations, is the only one which shows a sign patterns
for all four measured charge correlations that is qualitatively
similar to the STAR data.\footnote{One could imagine that tuning the strength
of each of the three components may provide a reasonable fit to the
presently available STAR data. Given the rather schematic correlations
employed in this study, the value of such an exercise is,
however, rather limited.}

\begin{table}[!h]
\caption{Various obervables in Test-I (see text).}\label{table_1}
\begin{tabular}{|c|c|c|c|c|}
    \hline
$\left<\hat O\right>$ &  IIIc    & IIIc-($\alpha$)  & IIIc-($\beta$) & IIIc-($\gamma$)   \\
    \hline
$\left<(Q^c_1)^2\right>$   %
& 799.8 & 437.8 & 437.6  & 253.1 \\
$\left<(Q_2)^2\right>$   %
& 1998.2 & 2024.8 & 2027.9  & 2047.0\\
$\left<(\Delta Q^2)\cdot Q_2\right>$   %
& -13109.4 & -7345.3 & -7400.1 & -4314.1 \\
    \hline
$\left<\cos(\phi_i-\phi_j)\right>_{++/--}$   %
& 0.0025 & -0.00078 & 0.00197  & -0.00131 \\
$\left<\cos(\phi_i-\phi_j)\right>_{+-}$   %
& -0.0025 & -0.00125 & 0.00149  & 0.00053 \\
$\left<\cos(2\phi_i-2\phi_j)\right>_{++/--}$   %
& 0.0100 & 0.0102 & 0.0102  & 0.0103 \\
$\left<\cos(2\phi_i-2\phi_j)\right>_{+-}$   %
& 0.0100 & 0.0102 & 0.0102  & 0.0103 \\
$\frac{1}{v_2}\left<\cos(\phi_i+\phi_j-2\phi_k)\right>_{++/--\, ,\, k-any}$   %
& -0.0025 & -0.00167 & -0.00107  & -0.00109 \\
$\frac{1}{v_2}\left<\cos(\phi_i+\phi_j-2\phi_k)\right>_{+-\, ,\, k-any}$   %
& 0.0025 & 0.00130 & 0.00192  & 0.00082 \\
 \hline
$\left<\cos(2\, \Delta\Psi)\right>$%
& -0.3118 & -0.2873 & -0.2889  & -0.2636 \\
   \hline
\end{tabular}
\end{table}

These results can be partly understood as follows: the physical
dipole separates positive and negative charges in the out-of-plane
direction, while both the SCBB and the OCSS correlations, either by
separating same-charge pairs or by focusing
opposite-charge pairs, tend to reduce the out-of-plane charge
separation.

\subsection{Test-II}

\begin{figure}[!h]
\begin{center}
\includegraphics[width=6.5cm]{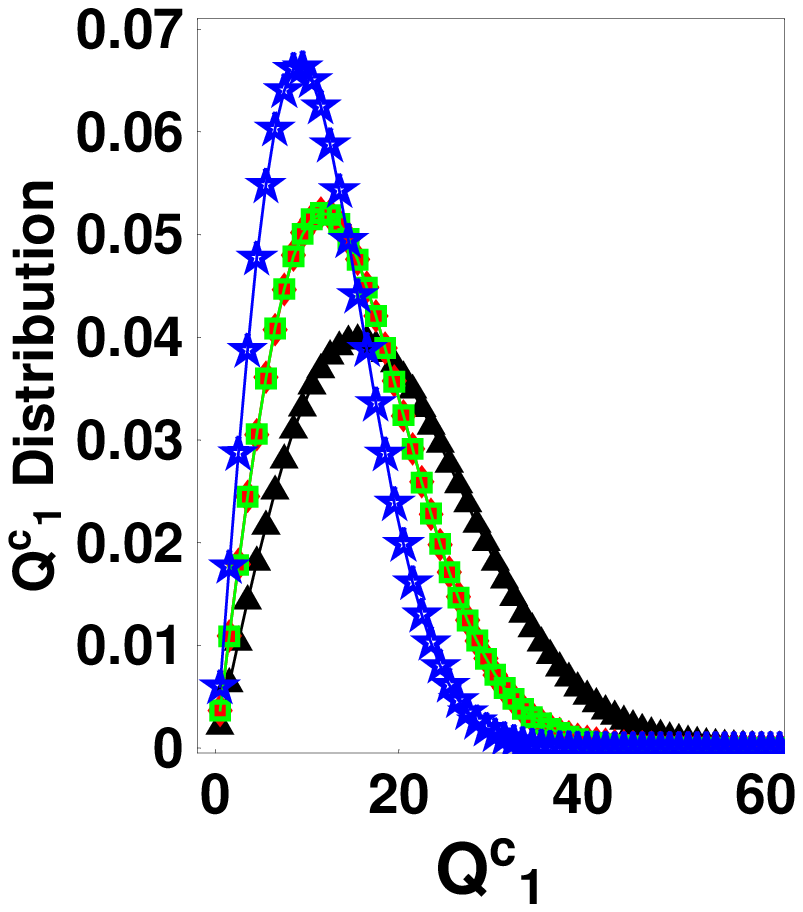}\hspace{1.cm}
\includegraphics[width=6.5cm]{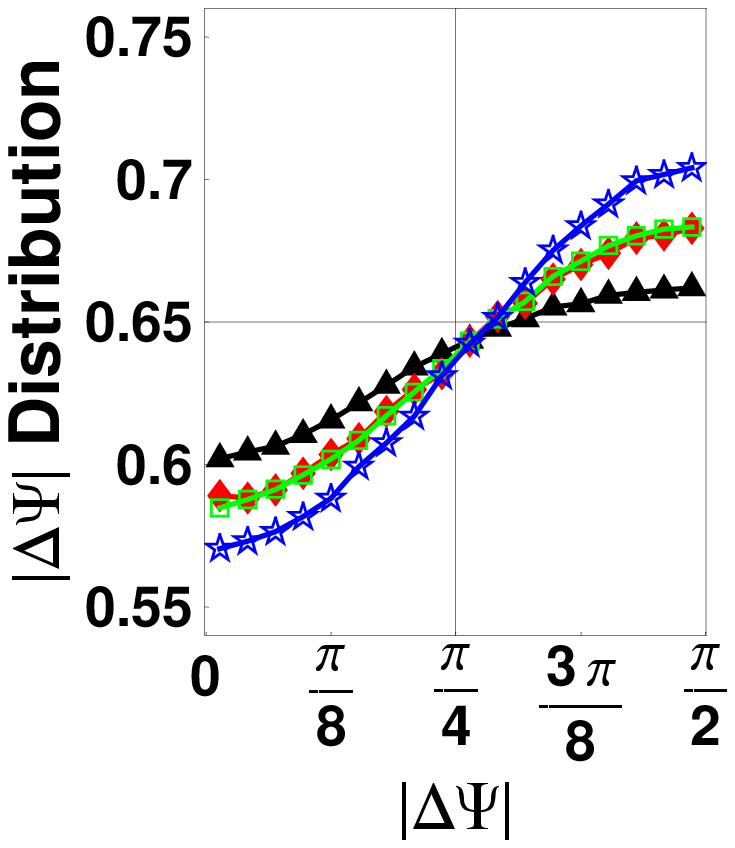}
\end{center}
\caption{The $Q^c_1$ (left) and $|\Delta \Psi|$  (right)
distributions for the four cases in Test-II. The triangle(black),
diamond(red), box(green), and star(blue) symbols are for Case-IV,
IV-($\alpha$), IV-($\beta$), and IV-($\gamma$) respectively (see
text for more details). } \label{fig_d02_corr}
\end{figure}

It is important to assess  how the above features change with the
magnitude of the built-in dipole. We thus also study a similar
setting to Test-I but with a smaller out-of-plane dipole
$d_1=0.02$. Again, we have four cases: {\bf Case-IV} with
the smaller dipole, and {\bf
Case-IV-($\alpha$),($\beta$),($\gamma$)} with the smaller dipole
plus the same correlations as in the {\bf
Case-IIIc-($\alpha$),($\beta$),($\gamma$)}. The results are shown
in Fig.\ref{fig_d02_corr} and Table.\ref{table_2}. The qualitative
conclusion
is largely the same as in Test-I, except for a few differences:\\
(1) Not surprisingly the angular distribution is much flatter due
to a smaller dipole, and the $\left<\cos(2\, \Delta\Psi)\right>$
is getting much closer to zero though still negative;\\
(2) {\bf Case-IV-($\beta$)} is interesting, since  one
finds
$\frac{1}{v_2}\left<\cos(\phi_i+\phi_j-2\phi_k)\right>_{++/--\,
,\, k-any}$ to be {\em positive} although there is still
out-of-plane charge separation (as indicated by angular
distribution and negative $\left<\cos(2\, \Delta\Psi)\right>$).

\subsection{Test-III}

\begin{figure}[!h]
\begin{center}
\includegraphics[width=6.5cm]{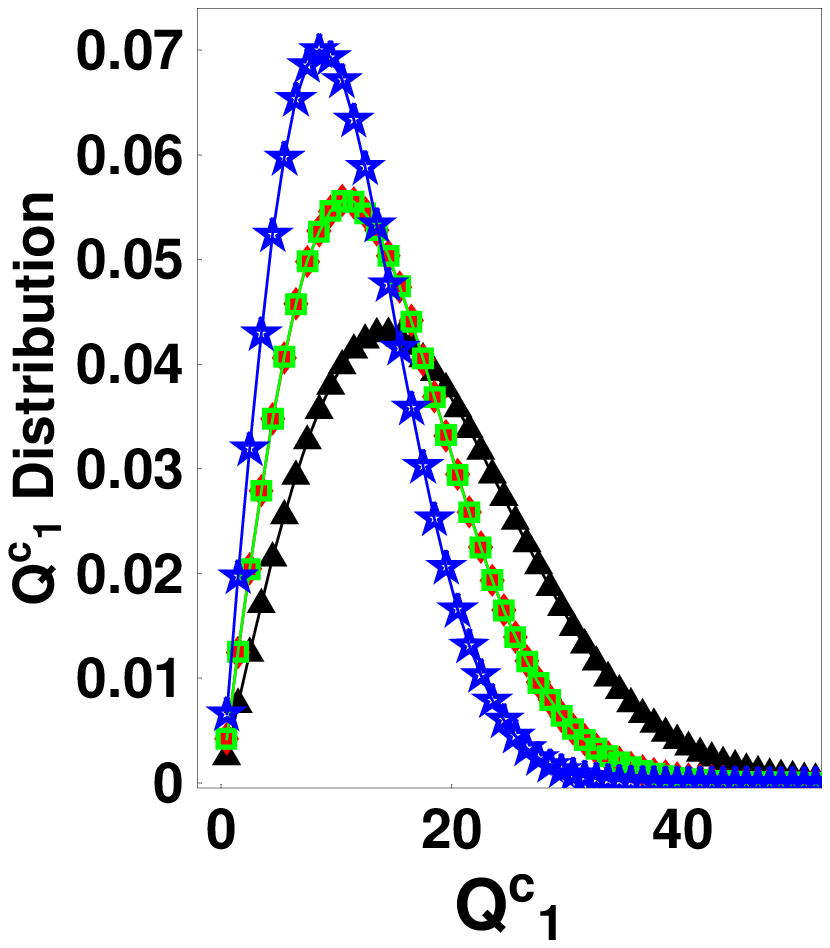}\hspace{1.cm}
\includegraphics[width=6.5cm]{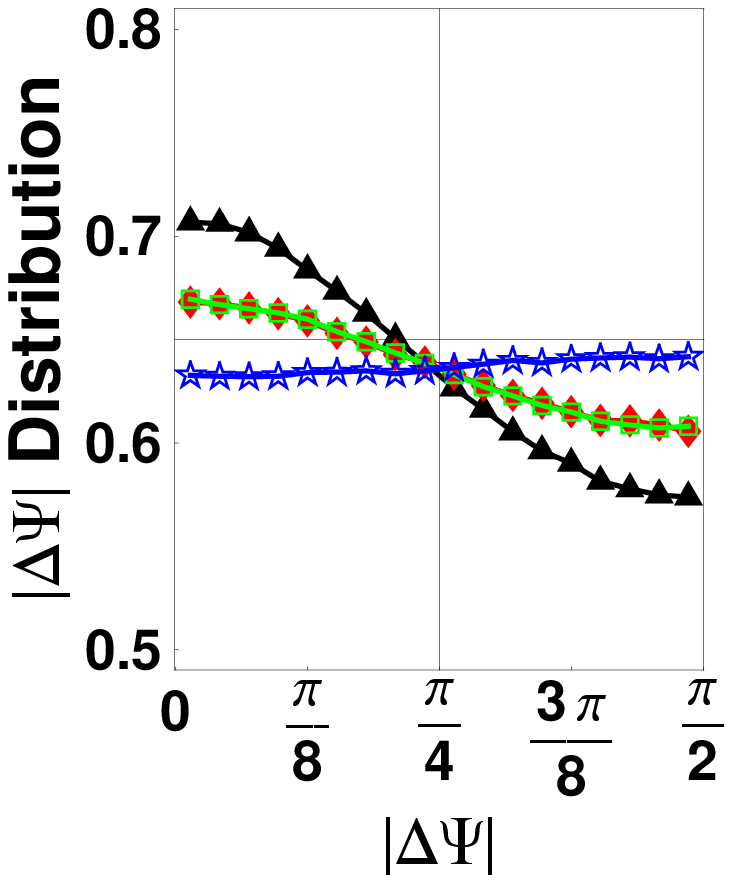}
\end{center}
\caption{The $Q^c_1$ (left) and $|\Delta \Psi|$  (right)
distributions for the four cases in Test-III. The triangle(black),
diamond(red), box(green), and star(blue) symbols are for Case-Ib,
Ib-($\alpha$), Ib-($\beta$), and Ib-($\gamma$) respectively (see
text for more details). } \label{fig_nod1_corr}
\end{figure}

In this test we completely turn off the dipole, i.e. $d_1=0$, in
order to expose the effects  due to correlations. Here the {\bf
Case-Ib} (previously studied) has elliptic flow but no dipole and
represents charge separation from pure statistical fluctuation,
while {\bf Case-Ib-($\alpha$),($\beta$),($\gamma$)} have no dipole
but the same correlations as in the {\bf
Case-IIIc-($\alpha$),($\beta$),($\gamma$)}. The results are
reported in Fig.\ref{fig_nod1_corr} and
Table.\ref{table_3}\footnote{In Table.\ref{table_3} and
\ref{table_4} certain extremely small values are given in the form
of $\hat{o}(10^{-n})$, which means they are consistent with zero
provided our finite statistics and may exactly vanish at infinite
statistics limit.}. A
comparison with Test-I,II shows the following:\\
(1) The correlator $\left<\cos(\phi_i-\phi_j)\right>$ for both same-charge
and opposite charge pairs show similar patterns as before, which
indicates that they are most sensitively dominated by  the
correlations;\\
(2) Most surprisingly, we find the correlations to rotate the
angular distribution in opposite direction to the previous
cases;\\
(3) Furthermore {\bf Case-Ib-($\gamma$)} (see last column of
Table.\ref{table_3}), even without a physical dipole,  shows a
sign pattern for all four measured charge correlations
qualitatively similar to the {\bf Case-IIIc-($\gamma$)} and to the
STAR data.

\begin{table}[!h]
\caption{Various obervables in Test-II (see text).}\label{table_2}
\begin{tabular}{|c|c|c|c|c|}
    \hline
$\left<\hat O\right>$ &  IV    & IV-($\alpha$)  & IV-($\beta$) & IV-($\gamma$)   \\
    \hline
$\left<(Q^c_1)^2\right>$   %
& 464.0 & 269.6 & 269.8  & 166.5 \\
$\left<(Q_2)^2\right>$   %
& 1996.5 & 2001.4 & 2001.7  & 2004.5\\
$\left<(\Delta Q^2)\cdot Q_2\right>$   %
& -423.8 & -704.1 & -716.5 & -719.3 \\
    \hline
$\left<\cos(\phi_i-\phi_j)\right>_{++/--}$   %
& 0.0004 & -0.00184 & 0.00091  & -0.00185 \\
$\left<\cos(\phi_i-\phi_j)\right>_{+-}$   %
& -0.0004 & -0.00020 & 0.00253  & 0.00107 \\
$\left<\cos(2\phi_i-2\phi_j)\right>_{++/--}$   %
& 0.0100 & 0.0100 & 0.0100  & 0.0101 \\
$\left<\cos(2\phi_i-2\phi_j)\right>_{+-}$   %
& 0.0100 & 0.0100 & 0.0100  & 0.0101 \\
$\frac{1}{v_2}\left<\cos(\phi_i+\phi_j-2\phi_k)\right>_{++/--\, ,\, k-any}$   %
& -0.0004 & -0.00056 & 0.00003  & -0.00049 \\
$\frac{1}{v_2}\left<\cos(\phi_i+\phi_j-2\phi_k)\right>_{+-\, ,\, k-any}$   %
& 0.0004 & 0.00021 & 0.00080  & 0.00022 \\
 \hline
$\left<\cos(2\, \Delta\Psi)\right>$%
& -0.0232 & -0.0375 & -0.0387  & -0.0527 \\
   \hline
\end{tabular}
\end{table}

\begin{table}[!h]
\caption{Various obervables in Test-III (see
text).}\label{table_3}
\begin{tabular}{|c|c|c|c|c|}
    \hline
$\left<\hat O\right>$ &  Ib    & Ib-($\alpha$)  & Ib-($\beta$) & Ib-($\gamma$)   \\
    \hline
$\left<(Q^c_1)^2\right>$   %
& 400.1 & 237.8 & 237.7  & 150.0 \\
$\left<(Q_2)^2\right>$   %
& 1996.7 & 1996.2 & 1995.2  & 1996.1\\
$\left<(\Delta Q^2)\cdot Q_2\right>$   %
& 1998.3 & 543.1 & 548.6 & -56.34 \\
    \hline
$\left<\cos(\phi_i-\phi_j)\right>_{++/--}$   %
& $\hat{o}(10^{-7})$ & -0.00204 & 0.00071  & -0.00196 \\
$\left<\cos(\phi_i-\phi_j)\right>_{+-}$   %
& $\hat{o}(10^{-7})$ &  $\hat{o}(10^{-7})$ & 0.00118  & 0.00053 \\
$\left<\cos(2\phi_i-2\phi_j)\right>_{++/--}$   %
& 0.0100 & 0.0100 & 0.0100  & 0.0100 \\
$\left<\cos(2\phi_i-2\phi_j)\right>_{+-}$   %
& 0.0100 & 0.0100 & 0.0100  & 0.0100 \\
$\frac{1}{v_2}\left<\cos(\phi_i+\phi_j-2\phi_k)\right>_{++/--\, ,\, k-any}$   %
& $\hat{o}(10^{-7})$ & -0.00036 & 0.00024  & -0.00038 \\
$\frac{1}{v_2}\left<\cos(\phi_i+\phi_j-2\phi_k)\right>_{+-\, ,\, k-any}$   %
& $\hat{o}(10^{-6})$ & $\hat{o}(10^{-7})$ & 0.00059  & 0.00011 \\
 \hline
$\left<\cos(2\, \Delta\Psi)\right>$%
& 0.0532 & 0.0242 & 0.0245  & -0.0038 \\
   \hline
\end{tabular}
\end{table}

It appears that on top of the charge separations due to
statistical fluctuations, the correlations studied here seem to
actually suppress the in-plane separation while ``enhancing'' the
out-of-plane separation to some extent. The reason for this behavior is
the presence of elliptic flow: while the two-particle correlations themselves
are reaction-plane independent, due to the elliptic flow
there are (on average) more pairs
close to the in-plane direction than  pairs close to the out-of-plane
direction. Therefore, the correlations are more effective in the
in-plane direction. Since, as already discussed, the correlations
considered here tend
to suppress  the charge separation, the
{\em statistical} charge separations are more suppressed
in-plane than out-of-plane. As a result we observe an effective
``enhancement'' of out-of-plane
charge separation. This subtle effect is proportional to the magnitude
of the elliptic flow, $v_2$,
and the correlation strength.

\subsection{Test-IV}

\begin{figure}[!h]
\begin{center}
\includegraphics[width=6.5cm]{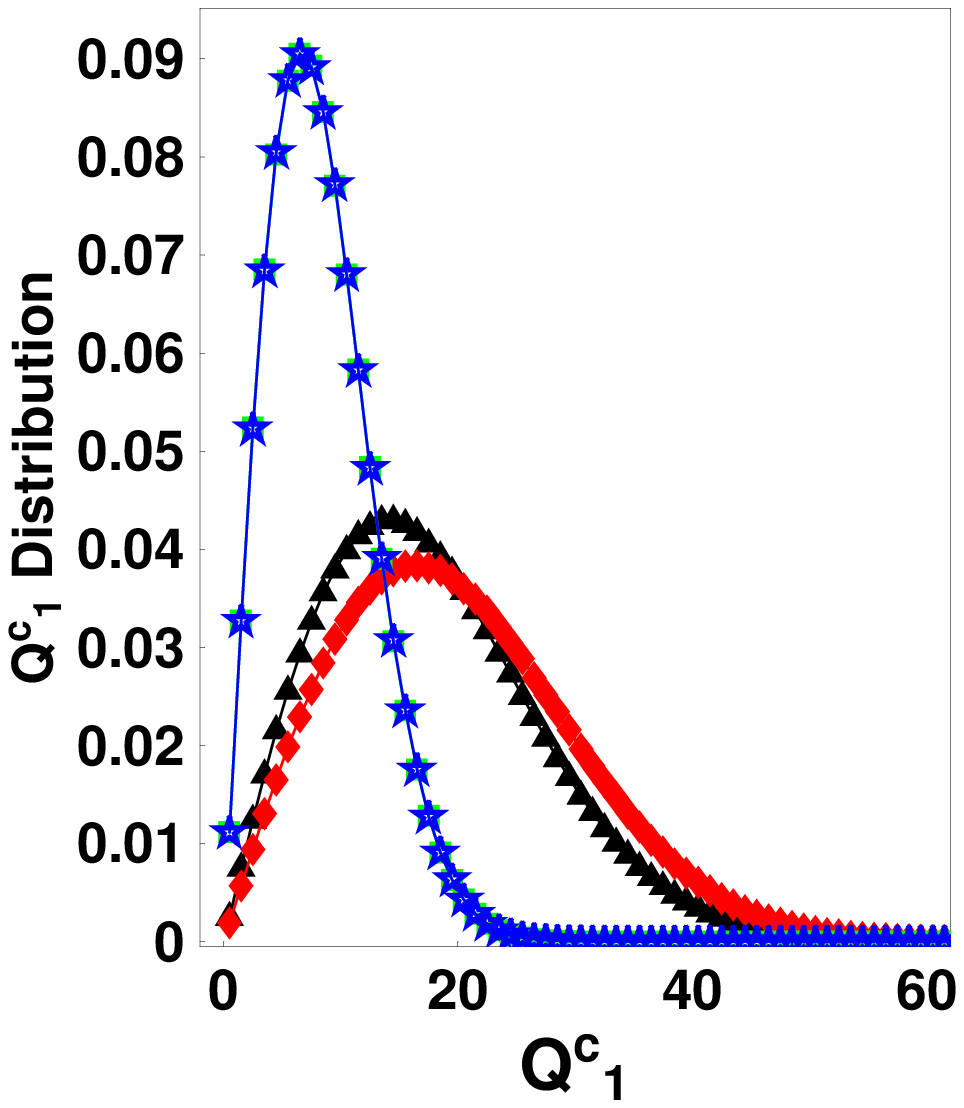}\hspace{1.cm}
\includegraphics[width=6.5cm]{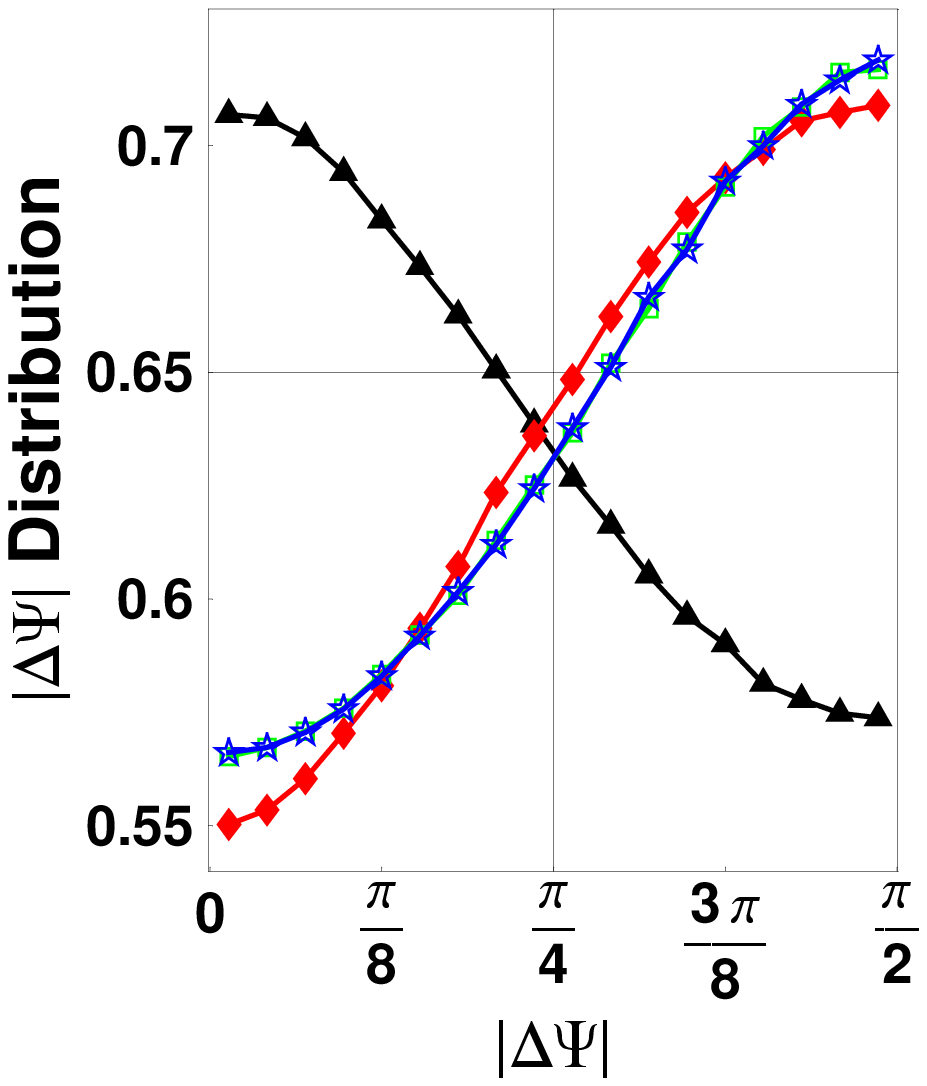}
\end{center}
\caption{The $Q^c_1$ (left) and $|\Delta \Psi|$  (right)
distributions for the four cases in Test-IV. The triangle(black),
diamond(red), box(green), and star(blue) symbols are for Case-Ib,
V-($\alpha$), V-($\beta$), and V-($\gamma$) respectively (see text
for more details). } \label{fig_contrast}
\end{figure}

The results in Test-III suggest an interesting question: could it
be that in a certain parameter region, the same out-of-plane charge
separation (e.g. the same $|\Delta\Psi|$ distribution) can be
produced either by a physical dipole (e.g. due to Chiral Magnetic
Effect) or by any one of the SCBB or OCSS correlations? It turns
out that this is  possible. Here in Test-IV we provide an example:\\
{\bf Case-V-($\alpha$)} with only a physical out-of-plane dipole $d_1=0.025$;\\
{\bf Case-V-($\beta$)} with no physical dipole and only SCBB
correlation at a fraction $f_{SCBB}=1\%$;\\
{\bf Case-V-($\gamma$)} with no physical dipole and only OCSS
correlation at a fraction $f_{OCSS}=1\%$.\\
In addition we also include the previously studied {\bf Case-Ib}
with only statistical fluctuation as a reference for comparison.
The results are presented in Fig.\ref{fig_contrast} and
Table.\ref{table_4}. A few very interesting features can be
immediately seen:\\
(1) All three cases V-($\alpha$)($\beta$)($\gamma$) show almost
identical $|\Delta\Psi|$ distribution and $\left<\cos(2\,
\Delta\Psi)\right>$, i.e. three distinctive physical effects can
lead to quantitatively similar ``apparent'' out-of-plane charge
separation;\\
(2) The physical dipole in V-($\alpha$), however, shifts the $Q^c_1$
distribution toward larger values while the correlations in
V-($\beta$)($\gamma$) shift the distribution toward smaller
values, thus making the $Q^c_1$ distribution also an important
discriminating probe;\\
(3) The observable
$\frac{1}{v_2}\left<\cos(\phi_i+\phi_j-2\phi_k)\right>$ has almost
the same value in V-($\alpha$) with physical dipole and in
V-($\beta$) with SCBB correlation and thus can not distinguish the
two cases, but it differentiates the V-($\gamma$) with OCSS
correlation;\\
(4) The correlator $\left<\cos(\phi_i-\phi_j)\right>$  again appears to be a
sensitive observable, giving different sign patterns for the three
cases.\\

\begin{table}[!h]
\caption{Various obervables in Test-IV (see text).}\label{table_4}
\begin{tabular}{|c|c|c|c|c|}
    \hline
$\left<\hat O\right>$ &  Ib    & V-($\alpha$)  & V-($\beta$) & V-($\gamma$)   \\
    \hline
$\left<(Q^c_1)^2\right>$   %
& 400.1 & 499.8 & 89.4  & 89.4 \\
$\left<(Q_2)^2\right>$   %
& 1996.7 & 1995.9 & 1996.2  & 1995.8\\
$\left<(\Delta Q^2)\cdot Q_2\right>$   %
& 1998.3 & -1762.9 & -499.4 & -497.2 \\
    \hline
$\left<\cos(\phi_i-\phi_j)\right>_{++/--}$   %
& $\hat{o}(10^{-7})$ & 0.00062 & -0.00390  & 0.00733 \\
$\left<\cos(\phi_i-\phi_j)\right>_{+-}$   %
& $\hat{o}(10^{-7})$ &  -0.00063 & $\hat{o}(10^{-7})$  & 0.01118 \\
$\left<\cos(2\phi_i-2\phi_j)\right>_{++/--}$   %
& 0.0100 & 0.0100 & 0.0100  & 0.0100 \\
$\left<\cos(2\phi_i-2\phi_j)\right>_{+-}$   %
& 0.0100 & 0.0100 & 0.0100  & 0.0100 \\
$\frac{1}{v_2}\left<\cos(\phi_i+\phi_j-2\phi_k)\right>_{++/--\, ,\, k-any}$   %
& $\hat{o}(10^{-7})$ & -0.00062 & -0.00059  & 0.00343 \\
$\frac{1}{v_2}\left<\cos(\phi_i+\phi_j-2\phi_k)\right>_{+-\, ,\, k-any}$   %
& $\hat{o}(10^{-6})$ & 0.00062 & $\hat{o}(10^{-8})$  & 0.00400 \\
 \hline
$\left<\cos(2\, \Delta\Psi)\right>$%
& 0.0532 & -0.06243 & -0.05996  & -0.05994 \\
   \hline
\end{tabular}
\end{table}

\newpage

The most important lessons we learn from these tests and
especially from Test-IV are:\\
first, the proposed $|\Delta \Psi|$ distribution and
$\left<\cos(2\,
\Delta\Psi)\right>$ can readily reveal the geometry of any potential charge separation;\\
second, an out-of-plane charge separation may arise due a physical
out-of-plane dipole (e.g. with Chiral Magnetic Effect) or due to
specific types of two-particle correlations (e.g. SCBB or OCSS);\\
third, the unambiguous way to disentangle these different sources
contributing to an out-of-plane charge separation, is to measure
all the available observables including both the $Q^c_1$ magnitude
and angular distribution and the STAR observables
$\frac{1}{v_2}\left<\cos(\phi_i+\phi_j-2\phi_k)\right>$ and
$\left<\cos(\phi_i-\phi_j)\right>$ as well.

\section{Summary}

To summarize, in this paper we have introduced and studied a method to
experimentally extract a possible charge separation in the
relativistic heavy ion collisions. Specifically we
have proposed to measure the event-by-event distribution of
the charged dipole vector, $\hat{Q}^c_1$.
This analysis is able to determine the
magnitude of the dipole as well as its azimuthal orientation with
respect to the reaction plane. Using Monte Carlo events, we have
investigated the sensitivity of this method for various scenarios,
including the presence of additional two-particle correlations.
We have shown that the combined information of the magnitude and
direction of the dipole can distinguish between effects due to
certain two particle correlations and those from e.g. the chiral magnetic
effect.

As in the case of the  elliptic flow analysis, the proposed method
may be refined by studying the $\hat{Q}^c_1$ distribution
differentially in transverse momentum or rapidity. Further
improvements may introduce transverse momentum or rapidity
dependent weight factors in the $\hat{Q}^c_1$ extraction via
Eq.(\ref{eqn_qc1_def}).

To conclude, we have demonstrated that our proposed method provides
additional discrimination power over already existing
measurements. This is essential in order to definitively determine the
presence of local parity violation in heavy ion collisions.

\bigskip

\section*{Appendix A}
\renewcommand{\theequation}{A\arabic{equation}}
\setcounter{equation}0

In this Appendix we show some details related to the charged
dipole vector $\hat{Q}^c_1$ analysis and discuss its relation to
various charged particle correlations.

The magnitude of $Q^c_1$ itself involves a square-root  and,
therefore,
contains all multi-particle correlations, as is
evident from a Taylor expansion:
\begin{eqnarray} \label{eqn:a1}
Q^c_1 &=& \sqrt{N_{ch}+\{i,j\}^c_1} = N_{ch}^{\frac{1}{2}} \left[ 1 + \frac{1}{2N_{ch}} %
\{i,j\}^c_1 - \frac{1}{8 N^2_{ch}} \left(\{i,j\}^c_1\right)^2 +
... \right]
\end{eqnarray}
with $\{i,j\}^c_1\equiv\sum_{i\ne j}q_i q_j \cos(\phi_i-\phi_j)$.
For example, the third term $\left(\{i,j\}^c_1\right)^2$ in the
above expansion involves 2,3,4-particle correlations. This can be
explicitly evaluated to give:
\begin{eqnarray} \label{eqn:a2}
\left(\{i,j\}^c_1\right)^2 = && N_{ch}(N_{ch}-1) \nonumber \\
&& + 2\sum_{i\ne j}\cos(\phi_i-\phi_j)+\sum_{i\ne
j}\cos2(\phi_i-\phi_j) \nonumber \\
&& + 2\sum_{i\ne j \ne k} q_i q_j \cos(\phi_i+\phi_j-2\phi_k)
\nonumber \\
&& + \sum_{i\ne j \ne k \ne l} q_i q_j q_k q_l
\cos(\phi_i+\phi_j-\phi_k-\phi_l)
\label{eq:a2}
\end{eqnarray}
We notice that the 3-particle correlation term in the above is
precisely the one proposed in \cite{Voloshin:2004vk} and measured
in \cite{Star:2009uh}. However, a full $\hat{Q}^c_1$ analysis
reveals more information than those expressed by the two and three
particle distributions. First, the expression for the magnitude of
$\hat{Q}_1^c$, Eq.(\ref{eqn:a1}), involves higher order terms than
Eq.(\ref{eqn:a2}). Thus it involves correlations beyond three
particles. Second, the determination of the magnitude
and azimuthal orientation of $\hat{Q}^c_1$
necessarily involves all  particles
in an event and hence correlations involving more than three
particles. Third, knowing the magnitude distribution of
$\hat{Q}^c_1$ allows to calculate the average of any moment
$<({Q}^c_1)^n>$, which contains n-particle correlations. As
discussed in \cite{Koch:2008ia}\cite{Bialas:1999tv} in general the
average of the $n-$th moment of an extensive observable like
$Q^c_1$ will necessarily involve  $n-$particle correlations.

As an example, from  the definition of  $\hat{Q}^c_1$,
Eq.(\ref{eqn_qc1_def}), one can calculate the $2^{nd}$ moment:
\begin{eqnarray} \label{eq:a3}
(Q^c_1)^2 &=& \sum_i q_i^2 (\cos^2\phi_i+\sin^2\phi_i) +
\sum_{i\ne j} q_i q_j (\cos\phi_i \cos\phi_j + \sin\phi_i
\sin\phi_j) \nonumber
\\
&=& N_{ch} + \sum_{i\ne j} q_i q_j \cos(\phi_i-\phi_j)
\end{eqnarray}
which is directly related to two particle azimuthal correlations
for both same-charge and opposite-charge pairs. This relation is
in analogy to the familiar relation for elliptic flow, i.e.
\begin{eqnarray} \label{eq:a4}
(Q_2)^2 &=& N_{ch} + \sum_{i\ne j} \cos(2\phi_i-2\phi_j)
\end{eqnarray}

Next we derive  Eq.(\ref{eqn_cos_2dPsi}). We first evaluate the
quantity
\begin{eqnarray}
(\Delta Q^2)\cdot Q_2 \equiv \left[(Q^c_1)^2 \cdot
\cos(2\Delta\Psi)\right]\cdot Q_2
\end{eqnarray}
by using the
definitions in Eq.(\ref{eqn_qc1_def},\ref{eqn_q2_def}):
\begin{eqnarray}
\left[(Q^c_1)^2 \cdot
\cos(2\Delta\Psi)\right]\cdot Q_2 &&= %
\left[(Q^c_1)^2\cos(2\Psi_1)\right]\cdot \left[Q_2\cos(2\Psi_2)\right]\nonumber \\
&&\quad
+\left[(Q^c_1)^2\sin(2\Psi_1)\right]\cdot \left[Q_2\sin(2\Psi_2)\right]\nonumber \\
&& =  N_{ch}  + 2\sum_{i\ne j}q_i q_j \cos(\phi_i-\phi_j)
\nonumber \\
&&\quad + \sum_{i\ne j}\cos2(\phi_i-\phi_j) + \sum_{i\ne j \ne k}
q_i q_j \cos(\phi_i+\phi_j-2\phi_k)
\end{eqnarray}
Combining the above with Eq.(\ref{eq:a3},\ref{eq:a4}), we obtain
\begin{eqnarray}
\cos(2\Delta\Psi)&&=\frac{\left[(Q^c_1)^2 \cdot
\cos(2\Delta\Psi)\right]\cdot Q_2}{(Q^c_1)^2 \, [(Q_2)^2]^{1/2}} \nonumber \\
&& = \frac{N_{ch}   + 2\, \{i,j\}^c_1 + \{i,j\}_2 +
\{i,j;k\}^c}{\left[N_{ch}+\{i,j\}^c_1\right]\cdot
\left[N_{ch}+\{i,j\}_2\right]^{1/2}}
\end{eqnarray}
with the two- and three- particle correlations $\{i,j\}^c_1,\,
\{i,j\}_2,\, \{i,j;k\}^c $ defined in Eq.(\ref{eqn_correlations}).
The Eq.(\ref{eqn_cos_2dPsi}) is simply the event averaged version
of the above expression.

\section*{Acknowledgements}
The authors are indebted to A. Poskanzer for very helpful
discussions. The authors also thank D. Kharzeev, R. Lacey, L.
McLerran, E. Shuryak, S. Voloshin, F. Wang, and N. Xu for
discussions and communications. This work was supported in part by
the Director, Office of Energy Research, Office of High Energy and
Nuclear Physics, Divisions of Nuclear Physics, of the U.S.
Department of Energy under Contract No. DE-AC02-05CH11231. A.B. is
also supported by the Polish Ministry of Science and Higher
Education, grant No. N202 125437 and the Foundation for Polish
Science (KOLUMB program).


\begin{thebibliography}{99}

\bibitem{'tHooft:1999au}
  G.~'t Hooft,
  arXiv:hep-th/0010225.

\bibitem{Schafer:1996wv}
  T.~Schafer and E.~V.~Shuryak,
  Rev.\ Mod.\ Phys.\  {\bf 70}, 323 (1998).

\bibitem{ES_book}
E.~Shuryak, {\it ``The QCD Vacuum, Hadrons and Superdense
Matter''}, 2nd ed., World Scientific Publishing Company, 2004.

\bibitem{Ripka:2003vv}
  G.~Ripka,
  arXiv:hep-ph/0310102.

\bibitem{Bali:1998de}
  G.~S.~Bali,
  arXiv:hep-ph/9809351.


\bibitem{Greensite:2003bk}
  J.~Greensite,
  Prog.\ Part.\ Nucl.\ Phys.\  {\bf 51}, 1 (2003).



\bibitem{Liao:2006ry}
  J.~Liao and E.~Shuryak,
  Phys.\ Rev.\  C {\bf 75}, 054907 (2007);
    Phys.\ Rev.\ Lett.\  {\bf 101}, 162302 (2008).
  M.~N.~Chernodub and V.~I.~Zakharov,
  Phys.\ Rev.\ Lett.\  {\bf 98}, 082002 (2007).
  A.~D'Alessandro and M.~D'Elia,
  Nucl.\ Phys.\  B {\bf 799}, 241 (2008).
  M.~Cristoforetti and E.~Shuryak,
  Phys.\ Rev.\  D {\bf 80}, 054013 (2009)
  [arXiv:0906.2019 [hep-ph]].
  A.~D'Alessandro, M.~D'Elia and E.~Shuryak,
  arXiv:1002.4161 [hep-lat].

\bibitem{Liao:2007mj}
  J.~Liao and E.~Shuryak,
  Phys.\ Rev.\  C {\bf 77}, 064905 (2008);
   Phys.\ Rev.\  D {\bf 73}, 014509 (2006);
  Nucl.\ Phys.\  A {\bf 775}, 224 (2006).

\bibitem{Shuryak:2008eq}
  E.~Shuryak,
  Prog.\ Part.\ Nucl.\ Phys.\  {\bf 62}, 48 (2009).
 D.~E.~Kharzeev,
  Nucl.\ Phys.\  A {\bf 827}, 118C (2009)
  [arXiv:0902.2749 [hep-ph]].


\bibitem{Liao:2008dk}
  J.~Liao and E.~Shuryak,
  Phys.\ Rev.\ Lett.\  {\bf 102}, 202302 (2009).
  E.~Shuryak,
  Phys.\ Rev.\  C {\bf 80}, 054908 (2009)
  [Erratum-ibid.\  C {\bf 80}, 069902 (2009)].
  C.~Ratti and E.~Shuryak,
  Phys.\ Rev.\  D {\bf 80}, 034004 (2009).
  M.~Lublinsky, C.~Ratti and E.~Shuryak,
  Phys.\ Rev.\  D {\bf 81}, 014008 (2010).
 D.~M.~Ostrovsky, G.~W.~Carter and E.~V.~Shuryak,
  Phys.\ Rev.\  D {\bf 66}, 036004 (2002).



\bibitem{Kharzeev:2004ey}
  D.~Kharzeev,
  Phys.\ Lett.\  B {\bf 633}, 260 (2006).

\bibitem{Kharzeev:2007tn}
  D.~Kharzeev and A.~Zhitnitsky,
  Nucl.\ Phys.\  A {\bf 797}, 67 (2007).

\bibitem{Kharzeev:2007jp}
  D.~E.~Kharzeev, L.~D.~McLerran and H.~J.~Warringa,
  Nucl.\ Phys.\  A {\bf 803}, 227 (2008).

\bibitem{Fukushima}
  K.~Fukushima, D.~E.~Kharzeev and H.~J.~Warringa,
  Phys.\ Rev.\  D {\bf 78}, 074033 (2008);
  arXiv:0912.2961 [hep-ph];
  arXiv:1002.2495 [hep-ph].
  D.~E.~Kharzeev and H.~J.~Warringa,
  Phys.\ Rev.\  D {\bf 80}, 034028 (2009).

\bibitem{Buividovich:2009wi}
  P.~V.~Buividovich, M.~N.~Chernodub, E.~V.~Luschevskaya and M.~I.~Polikarpov,
  Phys.\ Rev.\  D {\bf 80}, 054503 (2009).
 P.~V.~Buividovich, M.~N.~Chernodub, E.~V.~Luschevskaya and M.~I.~Polikarpov,
  Nucl.\ Phys.\  B {\bf 826}, 313 (2010).




\bibitem{Kharzeev:2009fn}
  D.~E.~Kharzeev,
  Annals Phys.\  {\bf 325}, 205 (2010).

\bibitem{Star:2009uh}
  B.~I.~Abelev {\it et al.}  [STAR Collaboration],
  Phys.\ Rev.\ Lett.\  {\bf 103}, 251601 (2009);
  arXiv:0909.1717 [nucl-ex].

\bibitem{Voloshin:2004vk}
  S.~A.~Voloshin,
  Phys.\ Rev.\  C {\bf 70}, 057901 (2004).


\bibitem{Bzdak:2009fc}
 A.~Bzdak, V.~Koch and J.~Liao,
  Phys.\ Rev.\  C {\bf 81}, 031901 (2010)
  [arXiv:0912.5050 [nucl-th]].

\bibitem{Wang:2009kd}
  F.~Wang,
  arXiv:0911.1482 [nucl-ex].


\bibitem{Pratt:2010gy}
  S.~Pratt,
  arXiv:1002.1758 [nucl-th].


\bibitem{Millo:2009ar}
  R.~Millo and E.~Shuryak,
  arXiv:0912.4894 [hep-ph].

\bibitem{Fukushima:2010vw}
  K.~Fukushima, D.~E.~Kharzeev and H.~J.~Warringa,
  arXiv:1002.2495 [hep-ph].

\bibitem{Basar:2010zd}
  G.~Basar, G.~V.~Dunne and D.~E.~Kharzeev,
  arXiv:1003.3464 [hep-ph].


\bibitem{Voloshin:2008dg}
  S.~A.~Voloshin, A.~M.~Poskanzer and R.~Snellings,
  arXiv:0809.2949 [nucl-ex].

\bibitem{Poskanzer:1998yz}
  A.~M.~Poskanzer and S.~A.~Voloshin,
  Phys.\ Rev.\  C {\bf 58}, 1671 (1998)
  [arXiv:nucl-ex/9805001].

\bibitem{Lacey}
R. Lacey, ``Local parity violation studies with PHENIX'', talk at
the ``Workshop on P- and CP-odd Effects in Hot and Dense Matter'',
http://www.bnl.gov/riken/hdm/.

\bibitem{Koch:2008ia}
  V.~Koch,
  arXiv:0810.2520 [nucl-th].


\bibitem{Bialas:1999tv}
  A.~Bialas and V.~Koch,
  Phys.\ Lett.\  B {\bf 456}, 1 (1999).



\end{thebibliography}
\end{document}